\documentclass[onecolumn,
 amsmath,amssymb,
 aps,
]{revtex4-2}

\usepackage{graphicx,xcolor}
\usepackage{dcolumn}
\usepackage{bm}

\begin{document}


\title{Pattern Formation in Thermal Convective Systems: Spatio-temporal Thermal Statistics, Emergent Flux, and Local Equilibrium}

\author{Atanu Chatterjee}
\email{atanu.chatterjee@weizmann.ac.il}
\affiliation{Weizmann Institute of Science, Department of Physics of Complex Systems, Rehovot, Israel} 
\author{Takahiko Ban}
\author{Atsushi Onizuka}
\affiliation{Osaka University, Division of Chemical Engineering, Department of Materials Engineering Science, Graduate School of Engineering Science, Machikaneyamacho 1-3, Toyonaka City, Japan}
\author{Germano Iannacchione} 
\affiliation{Worcester Polytechnic Institute, Department of Physics, Worcester, USA}


\begin{abstract}
We discuss spatio-temporal pattern formation in two separate thermal convective systems. In the first system, hydrothermal waves (HTW) are modeled numerically in an annular channel. A temperature difference is imposed across the channel, which induces a surface tension gradient on the free surface of the fluid, leading to a surface flow towards the cold side. The flow pattern is axially symmetric along the temperature gradient with an internal circulation for a small temperature difference. This axially symmetric flow (ASF) becomes unstable beyond a given temperature difference threshold, and subsequently, symmetry-breaking flow, i.e., rotational oscillating waves or HTW, appears. For the second system, Rayleigh-Bénard convection (RBC) is experimentally studied in the non-turbulent regime. When a thin film of liquid is heated, the competing forces of viscosity and buoyancy give rise to convective instabilities. This convective instability creates a spatio-temporal non-uniform temperature distribution on the surface of the fluid film. The surface temperature statistics are studied in both these systems as `order' and `disorder' phase separates. Although the mechanisms that give rise to convective instabilities are different in both cases, we find an agreement on the macroscopic nature of the thermal distributions in these emergent structures. 
\end{abstract}

\maketitle

\section{Introduction} 
Pattern formation during thermal convection is a well-studied phenomenon. The complex structures that emerge as a result of thermally driving the system out-of-equilibrium break spatio-temporal homogeneity. Generally, studies on pattern formation during thermal convection can be broadly grouped into two classes, namely, (i) when the system is close enough to equilibrium and (ii) when the system is driven far from equilibrium. While close to equilibrium phenomena offer exciting opportunities to study stable but complex structures and their slow relaxation to steady-states, far-from-equilibrium processes open up a plethora of questions regarding turbulence, dissipation, fast time scales, and large-scale order~\cite{cross1993pattern,jaeger2010far,bak1987self,nicolis1977self,prigogine1977time,georgiev2002least,georgiev2016road}. 

In this article, we focus on the near-equilibrium dynamics of two thermally driven systems: the Rayleigh-B{\'e}nard Convection (RBC) system and the Hydro-Thermal Wave (HTW) system. We confine ourselves to a pure thermodynamic study of these two systems; thus, the temperature is our key variable of interest. While the RBC system is experimentally studied in the non-turbulent regime (low Rayleigh number), the HTW system is explored over a broader domain, i.e., from axis-symmetric flows to complex rotational oscillating waves~\cite{chatterjee2019coexisting,chatterjee2019many,yadati2019spatio,ban2019thermodynamic,ban2020thermodynamic}. In the RBC system, thermal data of the top layer of the fluid film is obtained through infrared imaging; the HTW system, on the other hand, is simulated numerically based on the Navier-Stokes equations with appropriate boundary conditions. As these systems are driven out of equilibrium, the spatial symmetry is broken with the emergence of stable spatio-temporal complex patterns. While these complex patterns are recorded in-plane due to the global thermal driving, a thermodynamic flux orthogonal to the global driving force emerges due to the emergent complex thermal patterns~\cite{chatterjee2021evidence}. The thermal patterns and the emergent flux exist as long as the system is being driven. In the near-equilibrium regime, it is shown in this paper that the thermal statistics at the microscopic scale obtained from these two systems show similar behavior. Also, as a numerical model, the HTW system allows us to explore a larger parameter space than the RBC system, thus providing additional insights regarding possible far-from-equilibrium steady-states that may behave as attractors for these dissipative systems to asymptotically converge.

\section{Methodology}
It has been recently shown that RBC and HTW systems satisfy a linear force-flux relationship when close enough to equilibrium~\cite{ban2019thermodynamic,ban2020thermodynamic,chatterjee2021evidence}. Thus, under non-equilibrium conditions, the entropy production is maximized by the emergent thermodynamic flux, orthogonal to the global driving, in agreement with the maximum entropy production principle~\cite{onsager1931reciprocal,jou1996extended,lebon2008understanding,martyushev2006maximum}. While the linearity in the force-flux relationship concerns the bulk evolution of the system, it is imperative to ask if the underlying distribution of the local variables bears any key role on the nature of the macroscopic evolution of the system. The local variable of interest in the following two systems is temperature and its spatial distribution for our study. 

\subsection{Rayleigh-B{\'e}nard Convection}
\begin{figure}[t]
    \centering
    \includegraphics[scale=0.75]{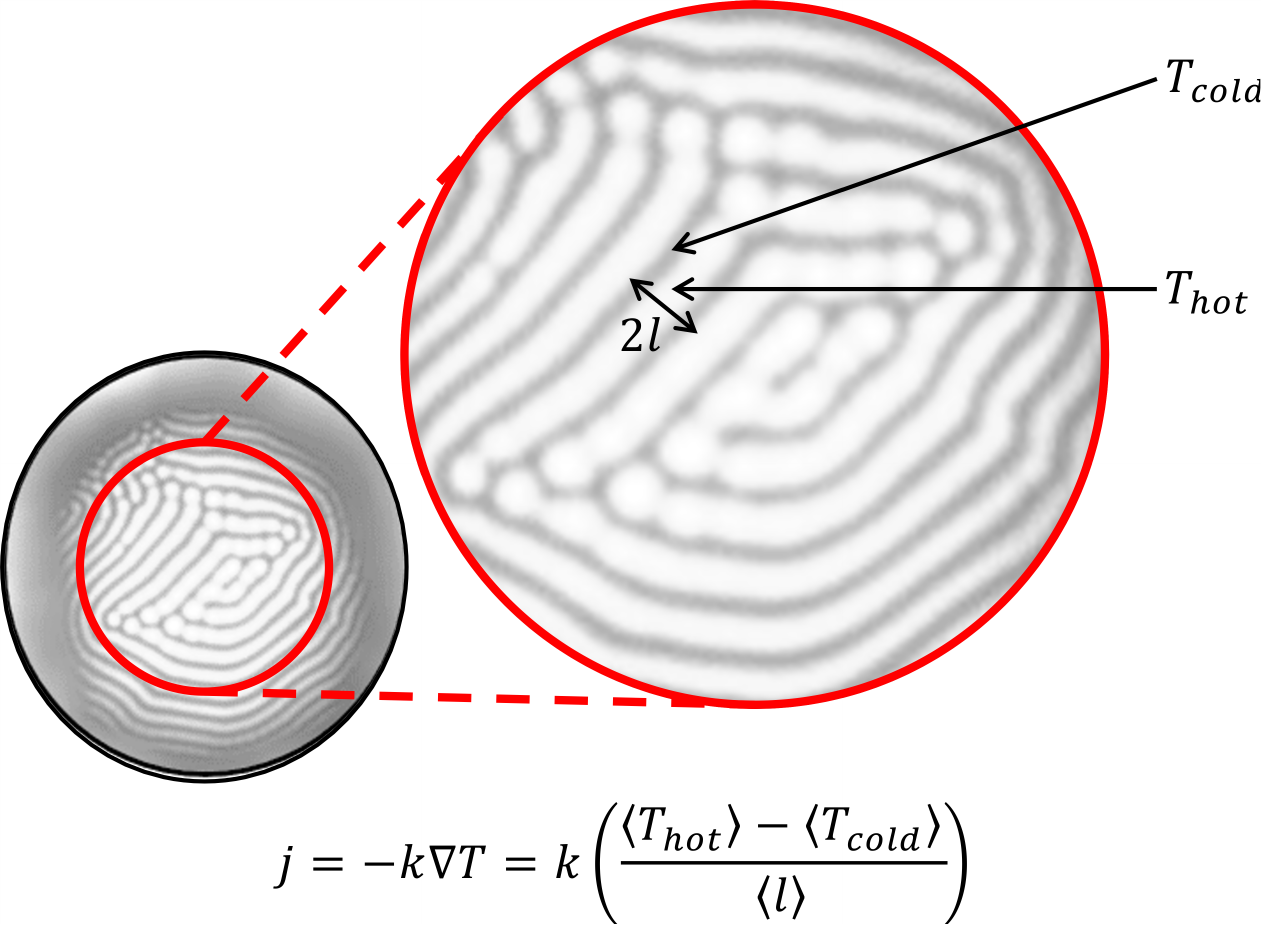}
    \caption{A steady-state RBC thermal image with striped patterns with a circular region of interest highlighted in red. Note that bright pixels are hotter than dark ones.}
    \label{fig1}
\end{figure}

In the RBC system, the temperature of the top fluid film is recorded using an infrared camera. A thin layer of high viscosity silicone oil is placed between a rigid-free boundary. It is driven from a room temperature equilibrium state to an out-of-equilibrium steady-state. As the system reaches a non-equilibrium steady-state, a fixed temperature difference is maintained between the free top layer and the rigid bottom layer, $T_{top}<T_{bottom}$. As the goal is to have convection cells over as wide as an area possible for the thermal imaging to yield significant temperature statistics, a large pan diameter to fluid-film thickness is chosen. A thermal dataset consisting of high-resolution grey-scale images is thus obtained by taking snapshots of the top layer of the fluid-film at regular intervals of $15$ seconds, capturing the moderate to slow dynamics of the emergence of the convection cell patterns as the system is being driven out-of-equilibrium. The thermal dataset allows us to obtain pixel-by-pixel temperature values, $T_k$, from the thermal scale present in each grey-scale image. The circular symmetry of the system is taken into account, and circular regions of interest are chosen. In Figure~\ref{fig1}, a snapshot of a thermal image from the RBC experiment is shown with the region of interest highlighted in red. First-order thermal statistics, such as mean, $\langle T\rangle$, and standard deviation, $\beta$ are then obtained over the pixels in the chosen region of interest as below,

\begin{equation}
    \langle T\rangle = \frac{1}{N}\sum_kT_k\quad\text{and}\quad\beta = \sqrt{\frac{1}{N-1}\sum_k\left(T_k-\langle T\rangle\right)^2}
    \label{eqn1}
\end{equation}

In a RBC system, the convective cells that emerge as a consequence of thermal driving is known to be three-dimensional. Since, the image snapshots capture only the top layer of the fluid film, our statistics are strictly restricted to the orthogonal plane that describe the top surface of the convective cells. Thus, the collection of in-plane hot ($\langle T_{hot}\rangle$) and cold ($\langle T_{hot}\rangle$) domains give rise to an emergent flux that is perpendicular to the global thermal driving force. The emergent flux is absent when there are no thermal patterns, even though an active thermal driving force is present. The mean separation between these domains as the system quenches to a non-equilibrium steady state is given by $\langle l\rangle$. With $k$ as the thermal conductivity of the working fluid, the emergent flux can be written as follows, 

\begin{equation}
     j = -k\nabla T = k\left(\frac{\langle T_{hot}\rangle - \langle T_{cold}\rangle}{\langle l\rangle}\right)
    \label{eqn2}
\end{equation}

The emergent flux denotes the onset of surface thermal gradients. It saturates once the system transitions to a steady-state. The growth and saturation of the emergent flux can be empirically described by the following differential equation,

\begin{equation}
     \tau\frac{dj(t)}{dt} + j(t) = j_\infty\quad\text{where}\quad j_{\infty} = k\left(\frac{\langle T_{hot}\rangle_\infty - \langle T_{cold}\rangle_\infty}{\langle l\rangle_\infty}\right)\quad\text{and}\quad j(t) = j_\infty\left(1 - e^{-t/\tau}\right)
    \label{eqn3}
\end{equation}

Here, $j_\infty$ denotes the steady-state value of the emergent heat-flux; therefore, $j(t)$ can also be expressed as $j(t) = k(\Delta T / l)_\infty (1 - e^{-t/\tau})$. The steady-state $\Delta T$ is the width of the steady-state bi-modal surface temperature distribution reported previously~\cite{chatterjee2019many,chatterjee2019coexisting,yadati2019spatio}. The exponent, $\tau$ is the time-constant of this lumped system, and has been used to quantify the Deborah number in the recent work~\cite{chatterjee2021evidence}.

\subsection{Hydro-Thermal Wave System}
In contrast, the HTW system is numerically solved in a three-dimensional setting ($r_{outer}=50~mm$, $r_{inner}=15~mm$, and $h=3~mm$) that involves an annular pool of silicon melt subjected to a fixed temperature difference between them, $T_{inner}<T_{outer}$. Specifically, the temperature of the inner wall is set at $T_{inner}=T_m=1683~K$ ($T_m$: melting point of silicon), and the temperature of the outer wall is varied in the range of $T_{outer}=\Delta T + T_{inner} = 1684-1697~K$. Adiabatic conditions were used for the upper surface and bottom surface, and the initial velocity of the fluid within the container was set to $0$ while the temperature was homogeneous at a value of $1683~K$.The governing equations are discretized using the finite volume method, and the numerical calculations are performed using OpenFoam, which uses the PISO algorithm. The number of grid points in the radial direction is chosen to be $81$, whereas $180$ in the circumferential direction, and $21$ in the vertical direction. These numbers for the grid are based on the conditions used by Li et al. and the resolution was sufficient to handle hydrothermal waves for a fluid with a low Prandtl number~\cite{li2004three,TAKAGI201472}. For the inner wall, outer wall, and the bottom surface of the container, rigid boundaries were used along with no-slip criteria for the velocity vector. At the free surface, the thermal Marangoni effect gives rise to convective flow. As the strength of natural convection relative to Marangoni convection is much weaker, the effect of gravity is ignored when solving the Navier-Stokes equation.

\begin{figure}[t]
    \centering
    \includegraphics[scale=0.75]{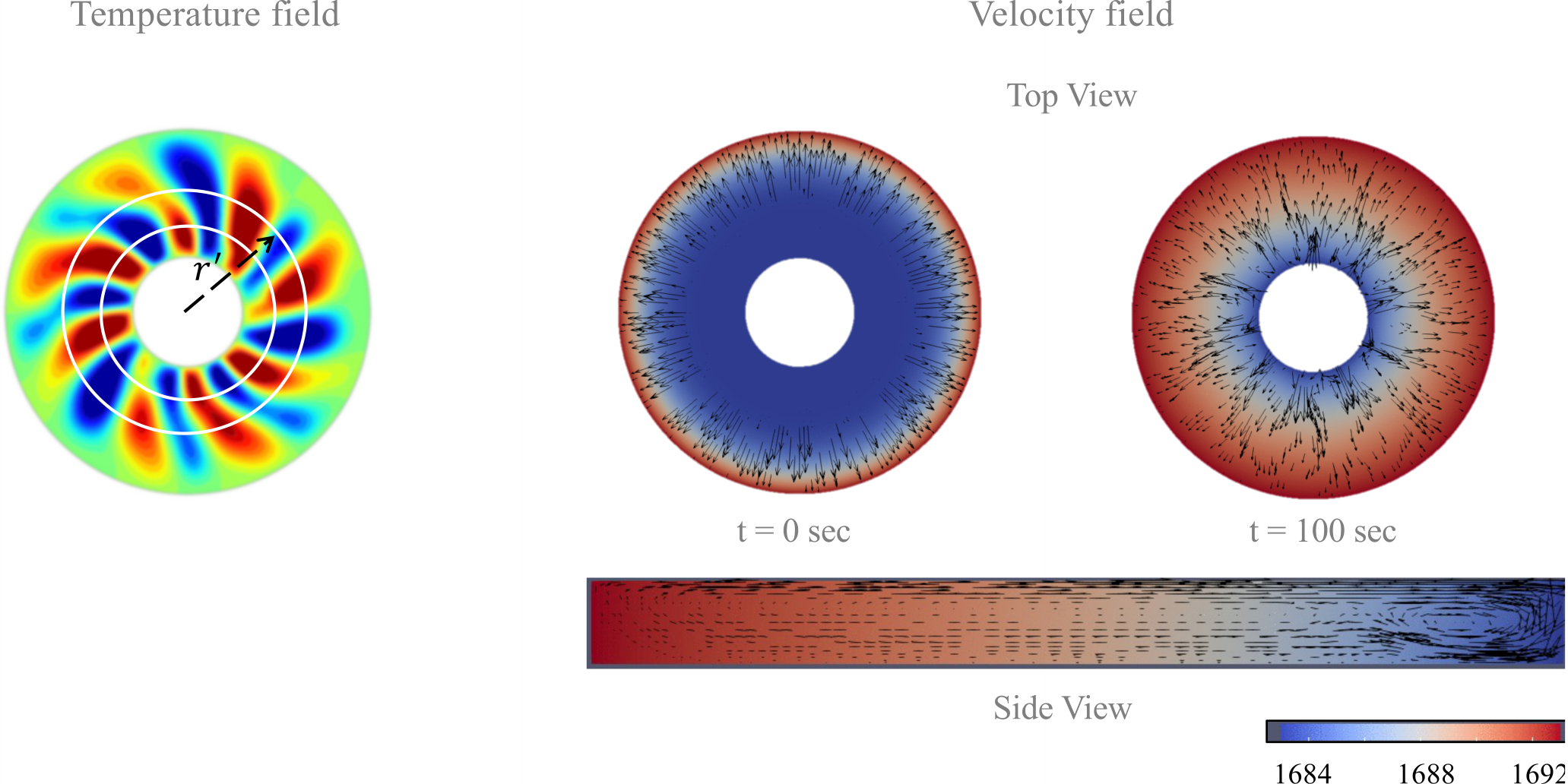}
    \caption{In the left panel, a steady-state HTW thermal image with rotating oscillatory waves is shown. The region of interest for the statistical analysis is denoted by the annular solid lines, and $r^\prime\sim20-30~mm$ is radially projected from the center of the setup. In the right panel, velocity flow fields of the HTW simulation are shown from the top at $t=0$ and $t=100$ seconds for a $\Delta T = 10~K$. A cross-sectional view of the flow lines is also shown at $t=100$ seconds.}
    \label{fig2}
\end{figure}

The temperature differential across the cell causes a surface tension gradient on the fluid's free surface, causing a surface flow to the cold side. As a result of the heat convection, numerous flow patterns with internal circulation emerge. The flow pattern in an annular channel is axially symmetric along the temperature gradient and includes internal circulation. Beyond a certain temperature difference threshold, this axially symmetric flow (ASF) becomes unstable, and symmetry-breaking flow, i.e., rotational oscillating waves, occurs (shown in Figure~\ref{fig2}). The oscillating waves propagate perpendicular to the temperature gradient applied to the system, i.e., the circumferential direction, and the temperature varies periodically. This circular oscillating flow is referred to as a hydro-thermal wave or HTW. 

In the HTW system, surface temperature statistics are obtained over a well-defined region of interest, as shown in Figure~\ref{fig2}. To account for the oscillatory waves, thermal statistics are also obtained along the angular coordinates such that, 

\begin{equation}
    \langle T\rangle_\theta = \frac{1}{2\pi}\int T(\theta)_{r=r^\prime}d\theta\quad\text{and}\quad\beta_\theta = \sqrt{\frac{1}{N-1}\sum_k\left(T(\theta_k)_{r=r^\prime}-\frac{1}{2\pi}\int T(\theta)_{r=r^\prime}d\theta\right)^2}
    \label{eqn4}
\end{equation}

Similarly, the emergent fluxes are obtained along both circumferential and radial directions. The emergent flux along the $\theta$ - direction is obtained by taking the angular gradient of the temperature scalar,

\begin{equation}
     j_\theta = -k\nabla_\theta T = -k
     \frac{1}{r^\prime}\frac{\partial T}{\partial\theta}
    \label{eqn5}
\end{equation}

The angular heat flux represents the magnitude of the interference effect between the heat conduction and viscous dissipation and is an energy source that produces emergent convection. The resulting emergent convection widens the angular temperature distribution. Thus, the angular standard deviation increases as the emergent convection grow. Therefore, the relationship between the normal standard deviation and the angular standard deviation may give the relationship between the applied energy and the energy required to form emergent convection. The HTW system is different from the RBC system because the emergent order in RBC quenches at a non-equilibrium steady-state. In the HTW system, the emergent order has a periodic behavior. Being a numerical simulation also allows us a wider parameter space to explore the various aspects of the stability of the oscillatory phenomena as a function of the input parameters and time.

\begin{figure}[t]
    \centering
    \includegraphics[scale=0.75]{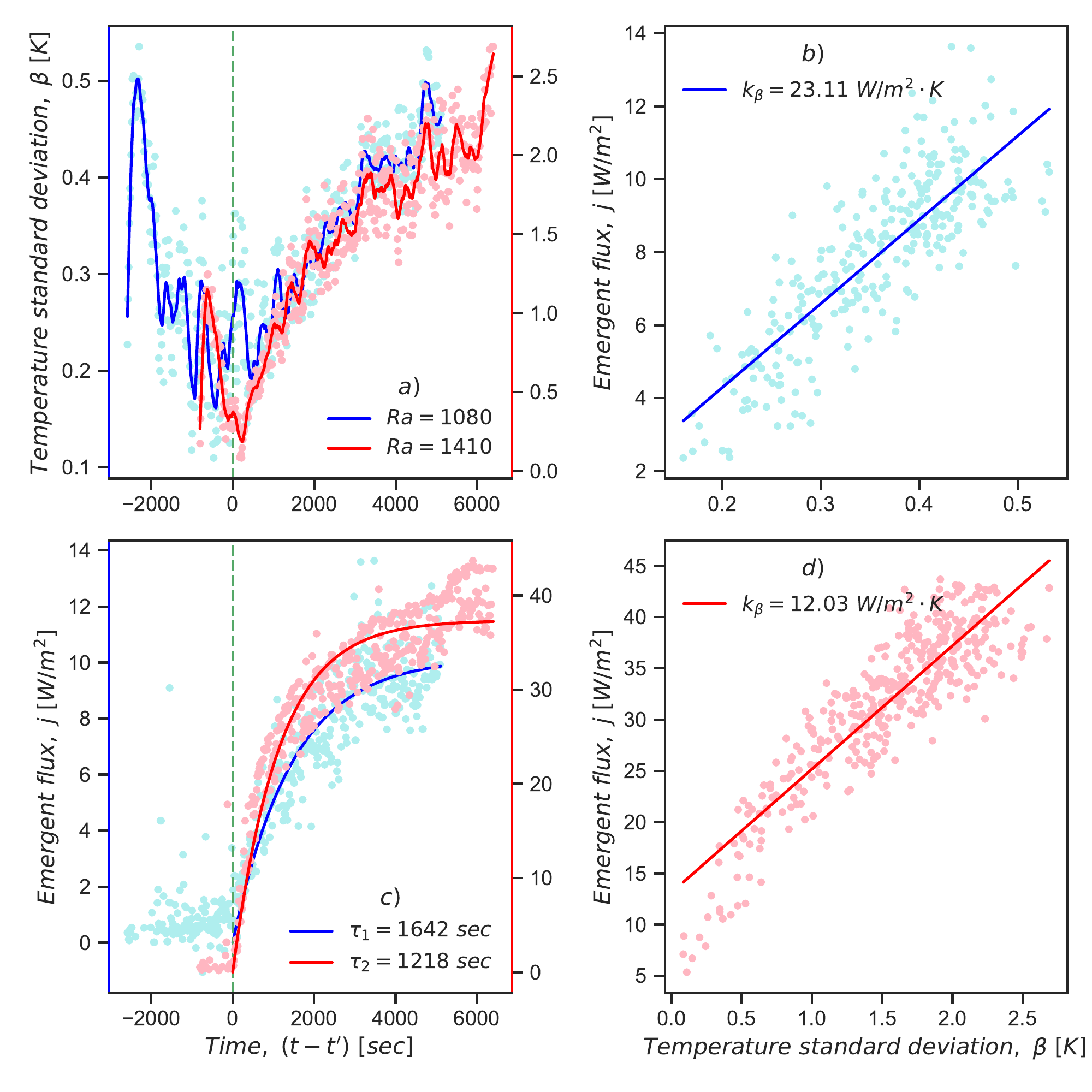}
    \caption{Time-series plot of the standard deviation, $\beta$ and emergent heat-flux, $j$ for the RBC system for two separate samples at $Ra=1080$ and $1410$ are shown in $a)$ and $c)$. In panel $c)$ the two solid lines show fits to the scatter data from Equation~\ref{eqn3} where $\tau$ denotes the two time-constants. Note that the time axis is shifted by $t^\prime$ which denotes the onset of surface thermal gradients (also denoted by the dashed vertical green line). Panels $b)$ and $d)$ portray the linear relationship between $j$ and $\beta$ once complex patterns emerge, i.e., for all the data points recorded at time, $t>t^\prime$.}
    \label{fig3}
\end{figure}

\section{Results}
In Figures~\ref{fig3}a and~\ref{fig3}c, we plot the temperature standard deviation and the emergent flux as a function of time for the RBC system. The different regions of the standard deviation time-series plot: growth, decline, and subsequent rise have been discussed in great detail in the previous works. Especially, plots discussing the stages of the emergent order as a function of the standard deviation or emergent flux for the RBC are not considered here as they have been previously reported~\cite{yadati2019spatio}. The solid lines in Figure~\ref{fig3}c denote the fits described in Equation~\ref{eqn3}. In Figures~\ref{fig3}b and~\ref{fig3}d, the linear relationship between emergent flux and surface temperature standard deviation are discussed for the two samples. The slopes of the linear fits ($k_\beta$) are also shown in the respective plots.

\begin{figure}[hb!]
    \centering
    \includegraphics[scale=0.15]{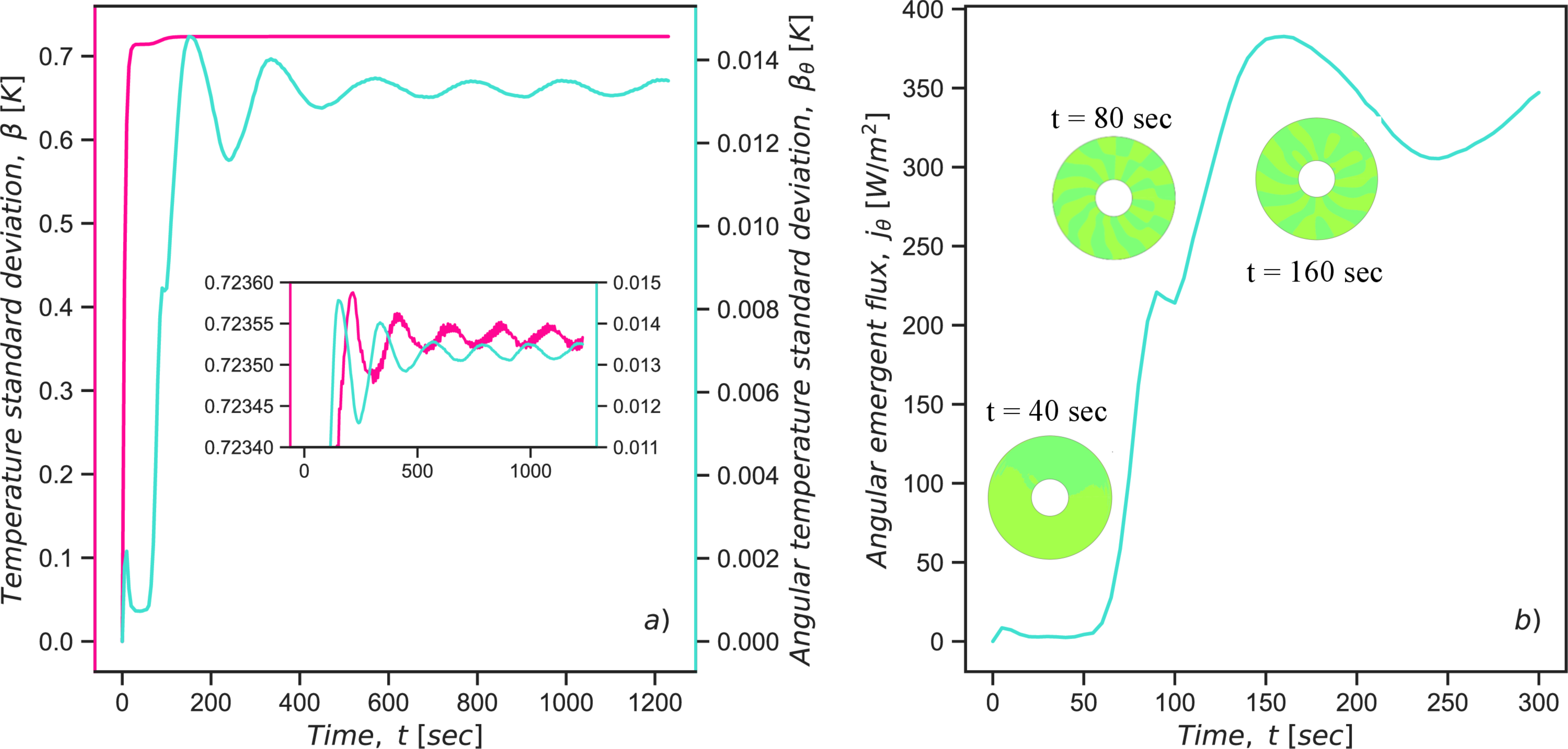}
    \caption{$a)$ Time-series plots of the surface temperature standard deviation, $\beta_\theta$ and $\beta$ for the HTW system are shown for a sample at $\Delta T = 7.5~K$. The inset plot shows the magnified view of the out of phase oscillations of $\beta$ and $\beta_\theta$. $b)$ Time-series plot of the angular emergent flux, $j_\theta$ as a function of time with snapshots of the top layer of the fluid at, $t=40,80$ and $160$ seconds.}
    \label{fig4}
\end{figure}

In this paper, we want to draw attention to the fact that in the HTW system as well, this trend in the surface temperature standard deviation time series is preserved. In Figure~\ref{fig4}, we present standard deviation and emergent flux plots calculated for the HTW system when the sample is subjected to a temperature difference, $\Delta T$ maintained at $7.5~K$. It is important to note that the two time-series plots in Figure~\ref{fig4}a, namely, the plots for the surface temperature standard deviation, oscillate out of phase with respect to each other. In Figure~\ref{fig4}b, snapshots of different stages of the emergent order are also shown at times, $t=40,80$ and $160$ seconds. Along the $\theta$-direction, it can be seen that the emergent flux and surface temperature standard deviation are in-phase. This may indicate a strong linear relationship between the two.   

\begin{figure}[t]
    \centering
    \includegraphics[scale=0.75]{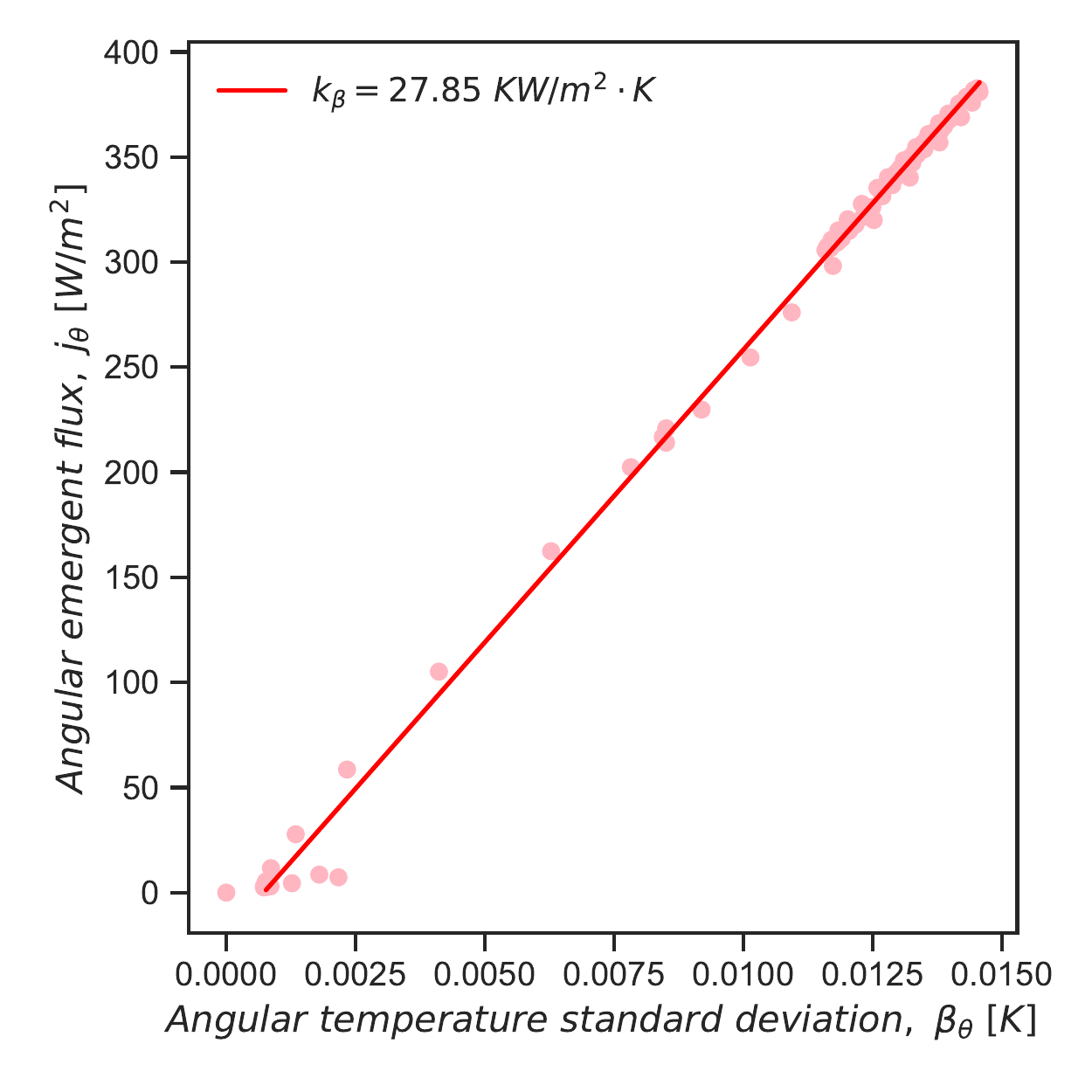}
    \caption{A linear relationship is observed between the standard deviation and emergent heat-flux for the HTW system for a sample at $\Delta T = 7.5~K$.}
    \label{fig5}
\end{figure}

In Figure~\ref{fig5}, standard deviation and emergent flux along $\theta$-direction are shown for the HTW simulations at $\Delta T = 7.5~K$. The relationship is observed to be linear after the onset of emergent order. It is interesting to note that before the onset of emergent order, the magnitude of $j_\theta$ remains very low, similar to the behavior observed in the RBC system (see Figure~\ref{fig3}b and~\ref{fig3}d), and has no correlation with surface temperature standard deviation, $\beta_\theta$. It is too early to comment whether there exists any connection between the two slopes obtained from the linear fits, $k_\beta$ in the RBC system, to that of the slope obtained from the linear fit for the HTW system. 

\begin{figure}[t]
    \centering
    \includegraphics[scale=0.6]{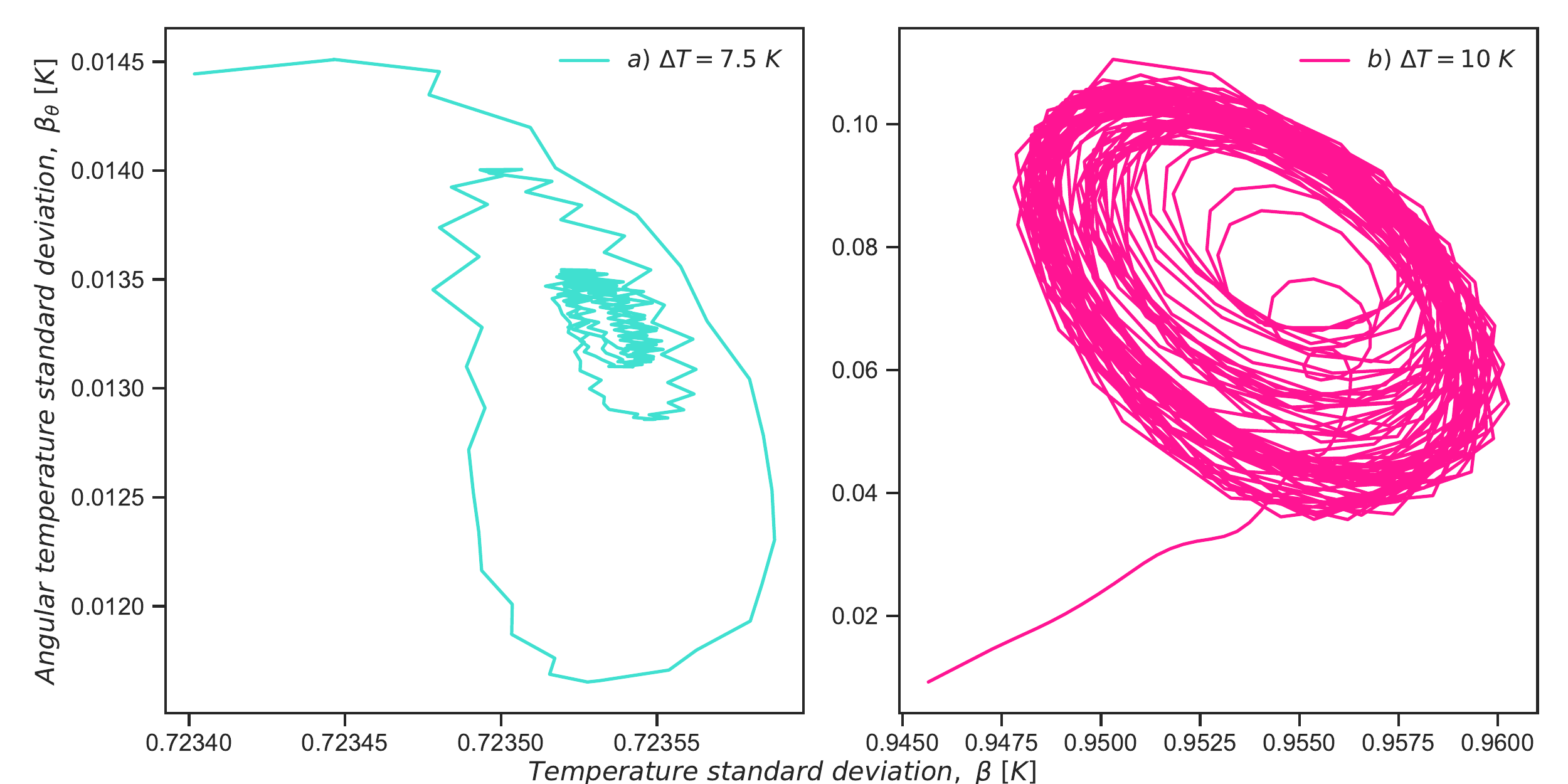}
    \caption{Limit cycle plots between the surface temperature standard deviation, $\beta$ and angular surface temperature standard deviation $\beta_\theta$ for different $\Delta T$ for the HTW system.}
    \label{fig6}
\end{figure}

Finally, in Figure~\ref{fig6}, we plot the two surface temperature standard deviation variables with respect to each other as the HTW system exhibits steady-state oscillations. We already discussed in Figure~\ref{fig4}a how these two variables oscillate out of phase with respect to each other. Therefore, in the $\beta-\beta_\theta$ phase-space these two variables give rise to limit cycles. As the system is dissipative and out-of-equilibrium, the arrows ($d\beta_\theta/d\beta < 0$) point to an attractor in phase-space that the system asymptotically evolves towards. 

\section{Discussion}
In this paper, we discuss pattern formation in two thermally driven convective fluid systems. We consider temperature as our thermodynamic variable of choice and use it to quantify spatio-temporal pattern formation and, in general, the macroscopic dynamics of the two systems under consideration. In the RBC system, the temperature is explicitly measured through infrared thermography. In the HTW system, the temperature at every lattice grid point is obtained by solving the Navier-Stokes equation. From the obtained temperature time-series data, first-order statistics are computed. While the time evolution of the mean temperature captures the macroscopic dynamics of the system as it relaxes to a non-equilibrium steady-state, the time evolution of the standard deviation is of particular interest as it allows us to identify the regimes in time when complex spatio-temporal patterns emerge. 

The peculiarity in the standard deviation time-series - steady rise followed by a steep drop and then a gradual rise -  as patterns emerge has been discussed before, both in the context of the RBC system and during pattern formation in other dissipative systems~\cite{de2019oscillatory,pal2019phase,pal2020concentration}. In this paper, we consider the HTW system numerically and observe similar peculiarity in the surface temperature standard deviation time-series plots, shown in Figure~\ref{fig4}. This further strengthens the argument against the observed unusual trend as an experimental artifact. Also, one can observe the similarity between the time evolution of the emergent flux and the surface temperature standard deviation, both in RBC and HTW systems from Figures~\ref{fig3} and~\ref{fig5}. Further, from Figures~\ref{fig3}b,~\ref{fig3}d and Figure~\ref{fig5} it can be established that in both systems, a linear relationship exists between emergent flux and surface temperature standard deviation after the onset of emergent order. In contrast, no correlation exists before order emerges. 

This may not be just a coincidence but may instead have deeper implications. For a low energy flux applied to the system (small temperature difference), the systems' response to the energy flux is symmetric, producing a uniform temperature distribution. In this case, ASF occurs in the (HTW) system, and the angular heat flux remains almost zero. The width of the temperature distribution increases with increasing the energy flux. Therefore, the standard deviation of the temperature distribution corresponds to the applied energy flux. Beyond the temperature difference threshold (or critical Rayleigh number, $Ra_c$), the symmetry of the temperature distribution is broken, and the heat flux that occurs perpendicular to the applied energy flux creates an emergent convection phenomenon in both HTW and RBC systems. The threshold is determined from the maximum entropy production principle~\cite{ban2019thermodynamic,ban2020thermodynamic}. The intersection of the entropy production curves of ASF and the HTW provides the transition point of the states. 

Recently, it has also been shown in the context of the local equilibrium hypothesis in RBC that the emergent flux and the thermodynamic force, calculated as the gradient of the inverse temperature, i.e., $X=\nabla(1/T)$ are linearly related~\cite{chatterjee2021evidence,taniguchi2007onsager,bedeaux1986nonequilibrium,bedeaux2003nonequilibrium}. For stability reasons, it has also been shown that at the near-equilibrium regime, the microscopic evolution of the patterns and the macroscopic evolution of the system are separated by time scales that are at least an order of magnitude apart~\cite{lavenda2019nonequilibrium,johannessen2003nonequilibrium,lebon2008understanding}. Therefore, a linear relationship between surface temperature standard deviation and emergent flux also implies a linear relationship between surface temperature standard deviation and thermodynamic force. We believe that this observation is non-trivial as this emergent thermodynamic force should comply with the second law of thermodynamics and decrease the potential of the system - as indicated by the largest decline during the time evolution of the surface temperature standard deviation, or equivalently lead to an increase in the entropy production during the dissipative process of pattern formation~\cite{prigogine1977time,martyushev2006maximum,chatterjee2016thermodynamics,clausius1854veranderte,chatterjee2019overview}.



\begin{thebibliography}{29}
\providecommand{\natexlab}[1]{#1}
\providecommand{\url}[1]{\texttt{#1}}
\expandafter\ifx\csname urlstyle\endcsname\relax
  \providecommand{\doi}[1]{doi: #1}\else
  \providecommand{\doi}{doi: \begingroup \urlstyle{rm}\Url}\fi

\bibitem[Cross and Hohenberg(1993)]{cross1993pattern}
Mark~C Cross and Pierre~C Hohenberg.
\newblock Pattern formation outside of equilibrium.
\newblock \emph{Reviews of modern physics}, 65\penalty0 (3):\penalty0 851,
  1993.

\bibitem[Jaeger and Liu(2010)]{jaeger2010far}
Heinrich~M Jaeger and Andrea~J Liu.
\newblock Far-from-equilibrium physics: An overview.
\newblock \emph{arXiv:1009.4874}, 2010.

\bibitem[Bak et~al.(1987)Bak, Tang, and Wiesenfeld]{bak1987self}
Per Bak, Chao Tang, and Kurt Wiesenfeld.
\newblock Self-organized criticality: An explanation of the 1/f noise.
\newblock \emph{Physical review letters}, 59\penalty0 (4):\penalty0 381, 1987.

\bibitem[Nicolis(1977)]{nicolis1977self}
Gregoire Nicolis.
\newblock Self-organization in nonequilibrium systems.
\newblock \emph{Dissipative Structures to Order through Fluctuations}, pages
  339--426, 1977.

\bibitem[Prigogine(1977)]{prigogine1977time}
Ilya Prigogine.
\newblock Time, structure and fluctuations.
\newblock \emph{Nobel Lectures in Chemistry 1971-1980}, pages 263--285, 1977.

\bibitem[Georgiev and Georgiev(2002)]{georgiev2002least}
Georgi Georgiev and Iskren Georgiev.
\newblock The least action and the metric of an organized system.
\newblock \emph{Open systems \& information dynamics}, 9\penalty0 (4):\penalty0
  371, 2002.

\bibitem[Georgiev and Chatterjee(2016)]{georgiev2016road}
Georgi~Yordanov Georgiev and Atanu Chatterjee.
\newblock The road to a measurable quantitative understanding of
  self-organization and evolution.
\newblock In \emph{Evolution and Transitions in Complexity}, pages 223--230.
  Springer, 2016.

\bibitem[Chatterjee et~al.(2019)Chatterjee, Yadati, Mears, and
  Iannacchione]{chatterjee2019coexisting}
Atanu Chatterjee, Yash Yadati, Nicholas Mears, and Germano Iannacchione.
\newblock Coexisting ordered states, local equilibrium-like domains, and broken
  ergodicity in a non-turbulent rayleigh-b{\'e}nard convection at steady-state.
\newblock \emph{Scientific reports}, 9\penalty0 (1):\penalty0 10615, 2019.

\bibitem[Chatterjee and Iannacchione(2019)]{chatterjee2019many}
Atanu Chatterjee and Germano Iannacchione.
\newblock The many faces of far-from-equilibrium thermodynamics: Deterministic
  chaos, randomness, or emergent order?
\newblock \emph{MRS Bulletin}, 44\penalty0 (2):\penalty0 130–133, 2019.
\newblock \doi{10.1557/mrs.2019.18}.

\bibitem[Yadati et~al.(2019)Yadati, Mears, and Chatterjee]{yadati2019spatio}
Yash Yadati, Nicholas Mears, and Atanu Chatterjee.
\newblock Spatio-temporal characterization of thermal fluctuations in a
  non-turbulent rayleigh--b{\'e}nard convection at steady state.
\newblock \emph{Physica A: Statistical Mechanics and its Applications}, page
  123867, 2019.

\bibitem[Ban and Shigeta(2019)]{ban2019thermodynamic}
Takahiko Ban and Keigo Shigeta.
\newblock Thermodynamic analysis of thermal convection based on entropy
  production.
\newblock \emph{Scientific reports}, 9\penalty0 (1):\penalty0 1--9, 2019.

\bibitem[Ban(2020)]{ban2020thermodynamic}
Takahiko Ban.
\newblock Thermodynamic analysis of bistability in rayleigh--b{\'e}nard
  convection.
\newblock \emph{Entropy}, 22\penalty0 (8):\penalty0 800, 2020.

\bibitem[Chatterjee et~al.(2021)Chatterjee, Ban, and
  Iannacchione]{chatterjee2021evidence}
Atanu Chatterjee, Takahiko Ban, and Germano Iannacchione.
\newblock Evidence of local equilibrium in a non-turbulent rayleigh-b{\'e}nard
  convection at steady-state.
\newblock \emph{arXiv preprint arXiv:2107.03678}, 2021.

\bibitem[Onsager(1931)]{onsager1931reciprocal}
Lars Onsager.
\newblock Reciprocal relations in irreversible processes. i.
\newblock \emph{Physical Review}, 37\penalty0 (4):\penalty0 405, 1931.

\bibitem[Jou et~al.(1996)Jou, Casas-V{\'a}zquez, and Lebon]{jou1996extended}
David Jou, Jos{\'e} Casas-V{\'a}zquez, and Georgy Lebon.
\newblock Extended irreversible thermodynamics.
\newblock \emph{Extended Irreversible Thermodynamics}, pages 41--74, 1996.

\bibitem[Lebon et~al.(2008)Lebon, Jou, and
  Casas-V{\'a}zquez]{lebon2008understanding}
Georgy Lebon, David Jou, and Jos{\'e} Casas-V{\'a}zquez.
\newblock \emph{Understanding non-equilibrium thermodynamics}, volume 295.
\newblock Springer, 2008.

\bibitem[Martyushev and Seleznev(2006)]{martyushev2006maximum}
Leonid~M Martyushev and Vladimir~D Seleznev.
\newblock Maximum entropy production principle in physics, chemistry and
  biology.
\newblock \emph{Physics reports}, 426\penalty0 (1):\penalty0 1--45, 2006.

\bibitem[Li et~al.(2004)Li, Imaishi, Azami, and Hibiya]{li2004three}
You-Rong Li, Nobuyuki Imaishi, Takeshi Azami, and Taketoshi Hibiya.
\newblock Three-dimensional oscillatory flow in a thin annular pool of silicon
  melt.
\newblock \emph{Journal of Crystal Growth}, 260\penalty0 (1-2):\penalty0
  28--42, 2004.

\bibitem[Takagi et~al.(2014)Takagi, Okano, Minakuchi, and Dost]{TAKAGI201472}
Y. Takagi and Y. Okano and H. Minakuchi and S. Dost.
\newblock Combined effect of crucible rotation and magnetic field on hydrothermal wave 
\newblock \emph{Journal of Crystal Growth}, 385\penalty0:\penalty0
  72--76, 2014.

\bibitem[De~Bari et~al.(2019)De~Bari, Dixon, Kay, and
  Kondepudi]{de2019oscillatory}
Benjamin De~Bari, James~A Dixon, Bruce~A Kay, and Dilip Kondepudi.
\newblock Oscillatory dynamics of an electrically driven dissipative structure.
\newblock \emph{PloS one}, 14\penalty0 (5):\penalty0 e0217305, 2019.

\bibitem[Pal et~al.(2019)Pal, Gope, Kafle, and Iannacchione]{pal2019phase}
Anusuya Pal, Amalesh Gope, Rumani Kafle, and Germano~S Iannacchione.
\newblock Phase separation of a nematic liquid crystal in the self-assembly of
  lysozyme in a drying aqueous solution drop.
\newblock \emph{MRS Communications}, 9\penalty0 (1):\penalty0 150--158, 2019.

\bibitem[Pal et~al.(2020)Pal, Gope, Obayemi, and
  Iannacchione]{pal2020concentration}
Anusuya Pal, Amalesh Gope, John~D Obayemi, and Germano~S Iannacchione.
\newblock Concentration-driven phase transition and self-assembly in drying
  droplets of diluting whole blood.
\newblock \emph{Scientific reports}, 10\penalty0 (1):\penalty0 1--12, 2020.

\bibitem[Taniguchi and Cohen(2007)]{taniguchi2007onsager}
Tooru Taniguchi and EGD Cohen.
\newblock Onsager-machlup theory for nonequilibrium steady states and
  fluctuation theorems.
\newblock \emph{Journal of Statistical Physics}, 126\penalty0 (1):\penalty0
  1--41, 2007.

\bibitem[Bedeaux(1986)]{bedeaux1986nonequilibrium}
D~Bedeaux.
\newblock Nonequilibrium thermodynamics and statistical physics of surfaces.
\newblock \emph{Adv. Chem. Phys}, 64:\penalty0 47--109, 1986.

\bibitem[Bedeaux et~al.(2003)Bedeaux, Johannessen, and
  R{\o}sjorde]{bedeaux2003nonequilibrium}
D~Bedeaux, E~Johannessen, and A~R{\o}sjorde.
\newblock The nonequilibrium van der waals square gradient model.(i). the model
  and its numerical solution.
\newblock \emph{Physica A: Statistical Mechanics and its Applications},
  330\penalty0 (3-4):\penalty0 329--353, 2003.

\bibitem[Lavenda(2019)]{lavenda2019nonequilibrium}
Bernard~H Lavenda.
\newblock \emph{Nonequilibrium statistical thermodynamics}.
\newblock Courier Dover Publications, 2019.

\bibitem[Johannessen and Bedeaux(2003)]{johannessen2003nonequilibrium}
E~Johannessen and D~Bedeaux.
\newblock The nonequilibrium van der waals square gradient model.(ii). local
  equilibrium of the gibbs surface.
\newblock \emph{Physica A: Statistical Mechanics and its Applications},
  330\penalty0 (3-4):\penalty0 354--372, 2003.

\bibitem[Chatterjee(2016)]{chatterjee2016thermodynamics}
Atanu Chatterjee.
\newblock Thermodynamics of action and organization in a system.
\newblock \emph{Complexity}, 21\penalty0 (S1):\penalty0 307--317, 2016.

\bibitem[Clausius(1854)]{clausius1854veranderte}
Rudolf Clausius.
\newblock {\"U}ber eine ver{\"a}nderte form des zweiten hauptsatzes der
  mechanischen w{\"a}rmetheorie.
\newblock \emph{Annalen der Physik}, 169\penalty0 (12):\penalty0 481--506,
  1854.

\bibitem[Chatterjee et~al.(2020)Chatterjee, Mears, Yadati, and
  Iannacchione]{chatterjee2019overview}
Atanu Chatterjee, Nicholas Mears, Yash Yadati, and Germano~S Iannacchione.
\newblock An overview of emergent order in far-from-equilibrium driven systems:
  From kuramoto oscillators to rayleigh--b{\'e}nard convection.
\newblock \emph{Entropy}, 22\penalty0 (5):\penalty0 561, 2020.

\end{thebibliography}

\end{document}